\documentclass[aps,pra,twocolumn,superscriptaddress]{revtex4}
\usepackage{graphicx}
\usepackage{bm}% bold math
\usepackage{amsmath}% equation formating
\usepackage[colorlinks=true,allcolors=blue]{hyperref}

\providecommand{\be}{\begin{equation}}
\providecommand{\ee}{\end{equation}}
\providecommand{\ba}{\begin{eqnarray}}
\providecommand{\ea}{\end{eqnarray}}

\usepackage{mathbbol}
\usepackage[makeroom]{cancel}

\begin{document}

\title{Enhancing quantum transport efficiency by tuning non-Markovian dephasing}

\author{S. V. Moreira}
\affiliation{Centro de Ci\^encias Naturais e Humanas, Universidade Federal do ABC - UFABC, Santo Andr\'e, Brazil}
\author{B. Marques}
\affiliation{Centro de Ci\^encias Naturais e Humanas, Universidade Federal do ABC - UFABC, Santo Andr\'e, Brazil}
\author{R. R. Paiva}
\affiliation{Centro de Ci\^encias Naturais e Humanas, Universidade Federal do ABC - UFABC, Santo Andr\'e, Brazil}
\author{L. S. Cruz}
\affiliation{Centro de Ci\^encias Naturais e Humanas, Universidade Federal do ABC - UFABC, Santo Andr\'e, Brazil}
\author{D. O. Soares-Pinto}
\affiliation{Instituto de F\'isica de S\~ao Carlos, Universidade de S\~ao Paulo, CP 369, 13560-970 S\~ao Carlos, S\~ao Paulo, Brazil}
\author{F. L. Semi\~ao}
\affiliation{Centro de Ci\^encias Naturais e Humanas, Universidade Federal do ABC - UFABC, Santo Andr\'e, Brazil}

\begin{abstract}
We consider the problem of energy transport in a chain of coupled dissipative quantum systems in the presence of non-Markovian dephasing. We use a model of non-Markovianity which is experimentally realizable in the context of controlled quantum systems. We show that non-Markovian dephasing can significantly enhance quantum transport, and we characterize this phenomenon in terms of internal coupling strengths of the chain for some chain lengths. Finally, we show that the phenomenon of dephasing-assisted quantum transport is also enhanced in the non-Markovian scenario when compared to the Markovian case. Our work brings together engineered environments, which are a reality in quantum technologies, and energy transport, which is typically discussed in terms of complex molecular systems. We then expect that it may motivate experimental work and further theoretical investigations on resources which can enhance transport efficiency in a controllable way. This can help in the design of quantum devices with lower dissipation rates, an important concern in any practical application.
\end{abstract}
\vskip2pc

\maketitle

\section{Introduction}

The investigation of new ways to overcome classical performance is a vast and challenging field of research. It is quite natural to think of genuinely nonclassical resources of quantum systems such as entanglement \citep{Horodecki}, coherence \citep{Baumgratz}, and invasiveness \cite{Moreira4} whenever one seeks to come up with ways to surpass the ability of classical systems to perform certain tasks. It is also interesting to investigate features of classical processes with a parallel in the quantum domain, such as non-Markovianity, to find routes to enhance the utility of quantum systems. 
This concept of Markovianity was originally defined for classical stochastic processes and means no memory in the sense that the past history of a stochastic variable is irrelevant to determine its future \citep{Budini}. All one needs to know is its value at the present time. In turn, non-Markovianity means the deviation from Markovian evolutions, which can be interpreted as a result of the persistence of memory effects  \citep{RevMarkov, Li, Breuer, deVega}. 
For quantum systems, non-Markovian evolutions are of central importance when there is a system-environment coupling, since closed systems are trivially governed by Markovian evolutions.

The phenomenon of energy transport is present in a wide variety of open systems. In particular, energy transfer caused by electronic coupling between molecular aggregates in photosynthetic complexes  \cite{PlenioTransport,Guzik} and polymeric samples \cite{poly} constitute important examples.  Indeed, much effort has been directed towards understanding how the environment impacts energy transport in coupled quantum systems. With a few exceptions, most studies have focused on Markovian models \citep{Schachenmayer, Semiao, Feist, PlenioTransport,Biggerstaff,EngVibAssEnTranfIons}, thereby not exploiting to the fullest the potentialities of possibly enhancing the efficiency of quantum transport through non-Markovianity. Exceptions include studies about vibrational coupling in the Fenna-Matthews-Olson complex \cite{fmo1,fmo2,fmo3}, a few examples in condensed matter physics where non-Markovian dynamics induced by fermionic environments has been studied \cite{fermi1,fermi2,fermi3}, and more recently in ion traps \cite{EnvoiAssTranspIonsMaier}. However, what seems to have not yet been properly exploited is the possibility of affecting energy transport in controlled quantum systems through carefully designed protocols to harness non-Markovianity \cite{Souza,paulo,fred,mauro, exp}. This would be in full agreement with recent applications of non-Markovianity in many quantum technological applications such as quantum key distribution \cite{nmqkd}, quantum metrology \cite{nmqm}, and quantum teleportation \cite{nmqt}.

Quantum dynamics is described by completely positive and trace-preserving (CPTP) linear maps upon which the concepts of Markovianity or non-Markovianity are developed. On the one hand, Markovianity means a CPTP dynamical map which causes continuous loss of information from the quantum open system to the environment \cite{BLP}. On the other hand, and in analogy with the classical case, Markovianity can be defined by considering CP divisibility \citep{RHP, Pollock, Pollock2, Milz, Buscemi}.  To be precise, let $\mathcal{E}_{(t_f,t_i)}$ be a CPTP dynamical map,  the action of which on density matrices at time $t_i$  leads to density matrices at time $t_f$ with $t_f\geq t_i$. According to this view, a Markovian dynamical process is that for which the map $\mathcal{E}_{(t_3,t_1)}$, from $t_1$ to $t_3$, can always be decomposed as $\mathcal{E}_{(t_3,t_1)}=\mathcal{E}_{(t_3,t_2)}\mathcal{E}_{(t_2, t_1)}$, with $\mathcal{E}_{(t_3,t_2)}$  completely positive for every $t_3\ge t_2 \ge t_1$  \citep{RHP}.

%Make a paragraph about the dephasing with the next
%\textcolor{blue}{Noise is always present in real quantum systems, but sometimes the inclusion of specific engineering noise can improve the performance, as in the dephasing-assisted transport schemes \cite{7}. Then, we also study the effect of time-dependent dephasing in the non-Markovian evolution. When the coherence of the system is affected by a time-dependent dephasing mechanism due to the system-environment coupling, it was shown \cite{Souza} that the Markovianity of the system is governed by the signal of the total time dephasing rate during the evolution. If at any instant of time the dephasing has a negative value the evolution is non-Markovian.}

In this work, we investigate the efficiency of energy transport in a linear chain of $N$ two-level dissipative systems where non-Markovian dephasing is induced and controlled by the introduction of ancillas which are locally coupled to each site of the chain, as described in \cite{Souza}. Such a mechanism has been employed in the experimental study of temporal correlations as indicators of non-Markovianity \cite{Souza}. We consider the transport of an excitation, initially in the first site of the chain, to a site $N+1$, which is also a two-level system and will be referred to as the sink site, to which energy is transferred irreversibly. In this way, the population of the sink site can be viewed as a figure of merit of the efficiency of the quantum transport. 
\section{The model}\label{Model}

\begin{figure}[h!]
\centering % para centralizarmos a figura
\includegraphics[scale=0.72]{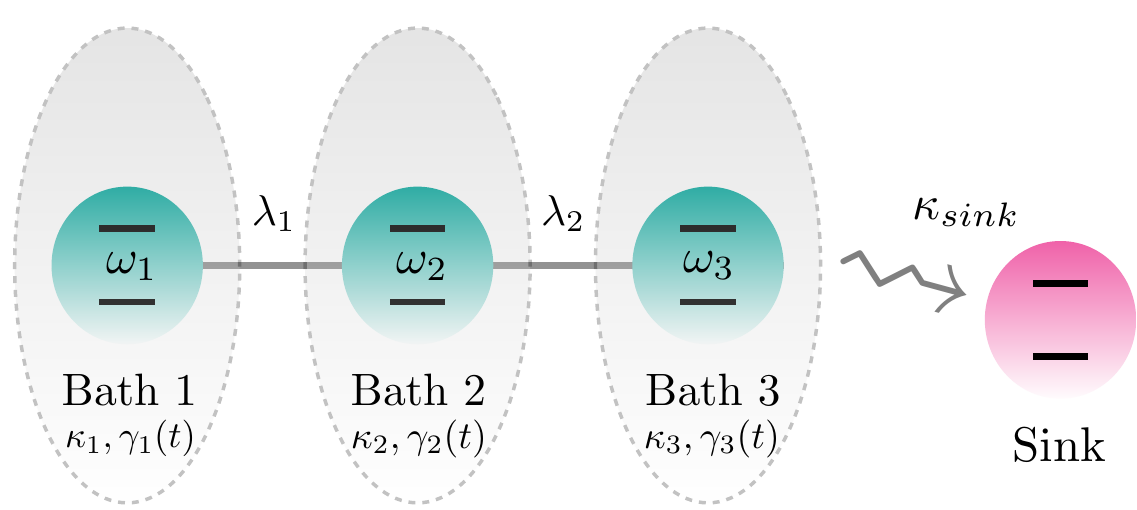}
\label{fig:fig14}
\caption{Illustration of a linear chain with three sites. Each site $i$ of the chain is coupled to its neighbor, as indicated by the solid orange line and the coupling strengths $\lambda_1$ and $\lambda_2$, as well as locally coupled to some environment which causes dissipation, with constant rate $\kappa_i$, and time-dependent dephasing at rate $\gamma_i(t)$. The last site of the chain can irreversibly give energy to a distinguished two-level system, named a sink, which can trap the excitation that is initially in the first site of the chain.} \label{ChainBath}
\end{figure}

We consider a chain with $N$-coupled two-level systems in a first-neighbor coupling model, implying a linear geometry, as represented in Fig. \ref{ChainBath}. The Hamiltonian that describes this system is given by ($\hbar=1$)
%We consider a chain with $N$ coupled two-level systems described by the Hamiltonian ($\hbar=1$)
\begin{equation}\label{H}
H = \sum_{i=1}^N \frac{\omega_i}{2}\sigma_i^z + \sum_{i=1}^{N-1} \lambda_i(\sigma_i^+\sigma_{i+1}^- + \sigma_{i+1}^+\sigma_i^- ),
\end{equation}
where $\sigma_i^+$ is the operator causing transition from the ground to the excited state in site $i$, $\sigma_i^-=(\sigma_i^+)^\dag$, $\sigma^z_i$ and $\omega_i$ are the Pauli $z$ operator and the energy associated with $i$th site, respectively, and $\lambda_i$ is the coupling constant between sites $i$ and $i+1$.  
%This represents a first-neighbor coupling model, implying a linear geometry for the transport of the excitations, as represented in Fig. \ref{ChainBath}.

Each site $i$ is subjected to local dissipation and local dephasing, such that the action of the superoperator $\mathcal{L}_i$ can be written as
%which account for these effects on the $i$th two-level system of the chain, can be written as \citep{Souza}
\begin{multline}\label{Li}
\mathcal{L}_i\rho = \kappa_i(2\sigma_i^-\rho\sigma_i^+-\sigma_i^+\sigma_i^-\rho-\rho\sigma_i^+\sigma_i^-) \\ +\gamma_i(t)(\sigma_i^z\rho\sigma_i^z-\rho)-is_i(t)[\sigma_i^z,\rho],
\end{multline}
where $\kappa_i$ are the damping rates responsible for energy dissipation, $\gamma_i(t)$ is a time-dependent dephasing rate, and $s_i(t)$ are environment-induced time-dependent energy shifts \citep{Souza}.
Here, the time evolution of the dissipation part will be assumed Markovian, which translates into $\kappa_i\ge0$.
Concerning the dephasing contribution, a Markovian evolution means $\gamma_i(t)\ge0$ and $s_i(t)\ge0$ for all times.

In the scope of controlled quantum systems, non-Markovian dephasing can be introduced and externally controlled \cite{Souza,paulo,fred,mauro, exp}. Usually one can achieve it by the introduction of controlled auxiliary systems or ancillas. Each ancilla interacts locally with each site in the chain and constitutes, with the original bath, the local environment affecting it. In this way, one can willingly make $\gamma_i(t)<0$ for some $i$ and for some time interval. This is well known to be a signature of non-Markovianity \cite{RevMarkov,BLP,RHP}, and this mechanism was experimentally demonstrated in the context of nuclear magnetic resonance  \citep{Souza} with the dephasing rate given by
\begin{equation}\label{gammat}
\gamma_i(t)= \gamma_i+ \frac{\pi J \sin^2(2\theta)\sin(2 \pi J t)}{3+2\cos(4\theta)\sin^2(\pi J t)+\cos(2 \pi J t )}.
\end{equation}
and an energy shift given by
\begin{equation}\label{st}
s_i(t)=\frac{2 \pi J \cos(2\theta)}{3+2\cos(4\theta)\sin^2(\pi J t)+\cos(2 \pi J t )},
\end{equation}
where $J$ and $\theta$ are fully controlled parameters. For a given set of uncontrolled environmental dephasing rates $\gamma_i\geq 0$, which depend on the physical subsystems forming the chain and their environment, it is possible to find and experimentally impose values of $J$ and $\theta$ which render $\gamma_i(t)\leq 0$ for some time $t$ \cite{Souza}. We will be using this non-Markovianity induction mechanism throughout this paper.

Finally, we consider that the $N$th site incoherently populates another two-level system which is usually named sink \cite{PlenioTransport}, as depicted in Fig. \ref{ChainBath}. Mathematically, this is described via the action of the superoperator $\mathcal{L}_{\text{sink}}$:
\begin{multline}\label{Lsink}
\mathcal{L}_{\text{sink}}\rho=\kappa_{\text{sink}}(2\sigma_{\text{sink}}^+\sigma_N^-\rho\sigma_N^+\sigma_{\text{sink}}^- \\- \sigma_N^+ \sigma_{\text{sink}}^-\sigma_{\text{sink}}^+\sigma_N^-\rho-\rho\sigma_N^+ \sigma_{\text{sink}}^-\sigma_{\text{sink}}^+\sigma_N^-),
\end{multline}
where $\kappa_{\text{sink}}\ge0$.
Notice that Eq. \eqref{Lsink} describes the irreversible transport of the excitation to the sink, which therefore traps it. Naturally, a relevant quantity for the study of quantum transport in such systems is the excitation transferred to the sink system, given by the population of the sink excited level at time $t$, $p_{\text{sink}}(t)={\rm Tr}(\rho(t)\sigma_{\text{sink}}^+\sigma_{\text{sink}}^-)$, where the system density matrix $\rho(t)$ obeys
\begin{equation}\label{Master}
\frac{\partial \rho(t)}{\partial t} =-i [H,\rho(t)] + \sum_{i=1}^N\mathcal{L}_i\rho(t)+\mathcal{L}_{\text{sink}}\rho(t).
\end{equation}
A figure of merit for the transport efficiency, $\eta$, is defined as $\lim_{t\to\infty} p_{\text{sink}}(t)$, which corresponds to the asymptotic value of the sink population. For the simulations,
the initial state, at $t=0$, is all sites and sink in their local ground states, except for the first chain-site which starts in the excited state.
 %we will always consider that, at $t=0$, the sink and all sites in the chain are in their local ground states, except for the first chain-site which starts in the excited state. 

To summarize, the excitation in the first site propagates through the chain always subjected to local dissipation and dephasing which can be made purely Markovian or non-Markovian, depending on the control parameters $J$ and $\theta$. The aim is to avoid the excitation to be trapped in the chain and to irreversibly populate the two-level sink system.
%The temporal evolution of the system's state, governed by \eqref{Master}, will be then considered for the followin  initial state of the system composed of the chain of $N$ spins plus the sink site: $\ket{\uparrow}_1\otimes\ket{\downarrow}_2\otimes \dots \otimes \ket{\downarrow}_N \otimes \ket{\downarrow}_{N+1}$.
%In the following, we study how non-Markovian dephasing affects the efficiency of transport $p_{\text{sink}}$ for different values of the couplings between the chain sites.

\section{Results and Discussion}\label{nonMarkovian}

\begin{figure}[h!]
\centering % para centralizarmos a figura
\includegraphics[width=8.5cm]{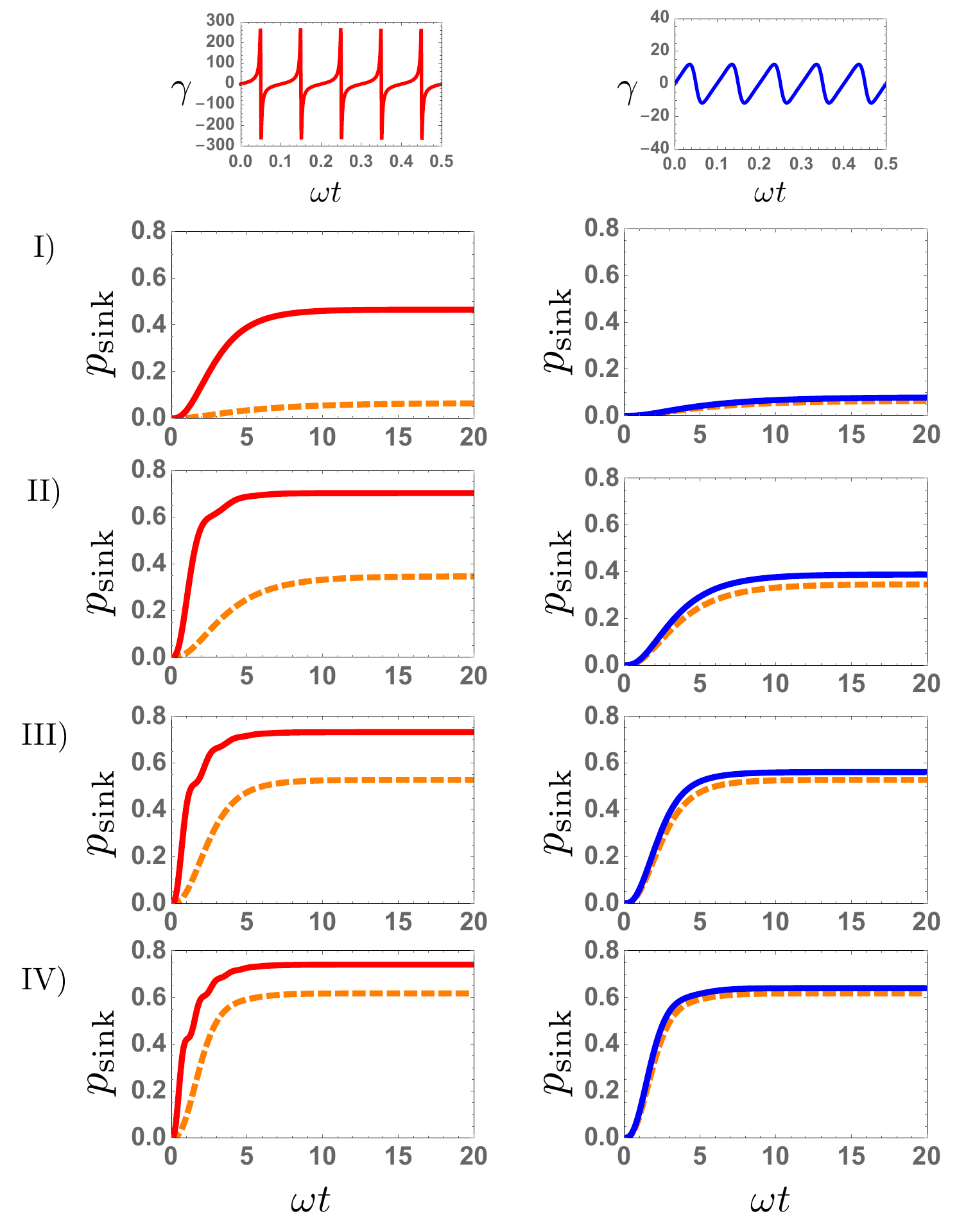}
\label{fig:fig14}
\caption{Plots of $p_{\text{sink}}$ as a function of $\omega t$ for $N=2$ and different values of the site-to-site coupling constant $\lambda$: I) $\lambda=0.1$ II) $\lambda=0.3$ III) $\lambda=0.5$ IV) $\lambda=0.7$. Orange dashed lines correspond to the Markovian case while red solid lines on the left correspond to $J=10$ and $\theta=0.8$, while  $J=10$ and $\theta=\pi/3$ are used for the blue solid lines on the right. Both of them correspond to non-Markovian evolutions, with the plots for the associated $\gamma(t)$  on the top. } \label{Coupling}
\end{figure}

We first consider the simplest scenario, a chain with two sites and the sink.
We solved Eq. \eqref{Master} numerically to determine $p_{\text{sink}}(t)$ for such a system. In Fig. \ref{Coupling}, we have some plots of $p_{\text{sink}}(t)$   for different values of $\lambda$, the coupling constant in the chain, and the Markovian and non-Markovian cases.  We fixed  $\kappa_1=\kappa_2=0.1$, $\gamma_1=\gamma_2=0.1$, and $\kappa_{\text{sink}}=0.6$ for all plots in Fig. \ref{Coupling}.  These numbers should be understood as in units of $\omega_1=\omega_2=\omega$.

%Time is then expressed in arbitrary units (a.u.), i.e., in the unit of the inverse of the frequencies of the problem (kept here unspecified).

For $N=2$, we numerically checked that as long as $\theta$ and $J$ in Eq. \eqref{gammat} lead to $\gamma(t)\geq 0$ for all $t$, the efficiency $\eta$, i.e. the asymptotic value of $p_{\text{sink}}$, is approximately the same. Consequently, without any loss of generality, we will set $J=0$ as the Markovian benchmark. For the non-Markovian case, $p_{\text{sink}}(t)$ is represented by the solid red lines on the left, with $(J=10,\theta=0.8)$, and solid blue lines on the right, with $(J=10,\theta=\pi/3)$. Both sets of values lead to negative values of the dephasing rate $\gamma(t)$; see the plots in Fig. \ref{Coupling}.

As one can see from the plots on the right in Fig. \ref{Coupling}, the presence of non-Markovian dephasing leads to some small enhancement of the transport efficiency.
%As one can see in Fig. \ref{Coupling}, the presence of non-Markovian dephasing leads to some small enhancement of the transport efficiency, as shown in the plots on the right.
Nonetheless, in the plots on the left, we can see an impressive enhancement.
In this way, under appropriate conditions, non-Markovianity may be used as a tool to enhance quantum transport in controlled quantum systems. Interestingly, by closely inspecting the plots in Fig. \ref{Coupling}, one also sees that the enhancement in the efficiency due to non-Markovianity tends to vanish for larger values of $\lambda$. The physical picture behind this fact is that strongly coupled sites lead to delocalized energy states or excitons. Consequently, non-Markovianity as given by Eq. \eqref{Master} tend to have its relevance reduced due to its localized action.

\begin{figure}[h!]
\centering % para centralizarmos a figura
\includegraphics[scale=0.30]{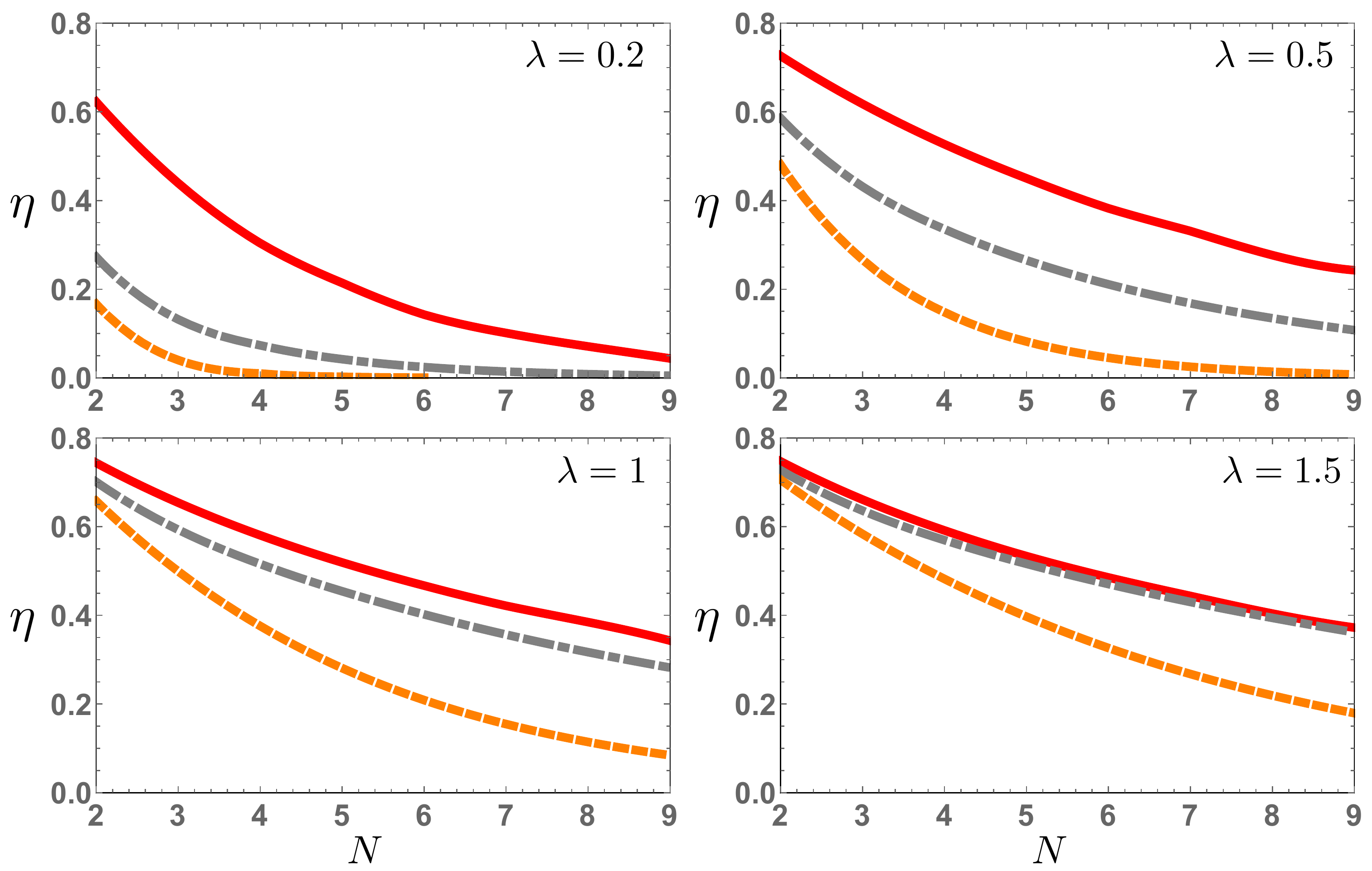}
\label{fig:fig14}
\caption{Plots of the transport efficiency $\eta$ as a function of the number of sites $N$ of the chain and different values of the site-to-site coupling constant $\lambda$, indicated in the plots. Orange dashed lines correspond to the Markovian case, while red solid lines on the left correspond to $J=10$ and $\theta=0.8$. The dashed-dotted gray lines represent the absence of dephasing, i.e., $\gamma_i(t)=0$ for all sites $i$.}\label{Coupling23}
\end{figure}

In Fig. \ref{Coupling23} we present plots of the transport efficiency $\eta$ as a function of the number of sites of the chain, $N$, for different values of the internal coupling strength $\lambda$. Once again we choose $J=0$ for the Markovian case, which corresponds to the orange dashed line. To obtain non-Markovian dephasing, $J=10$ and $\theta=0.8$ are set, just like in Fig. \ref{Coupling}. For the other parameters, $\omega_i=2$, $\kappa_i=0.1$, $\gamma_i=0.2$,  and $\kappa_{\text{sink}}=0.6$.
The non-Markovian case corresponds to the red solid line.
Here, we also consider the transport efficiency in the absence of dephasing,  i.e., $\gamma_i(t)=0$ for all sites $i$ of the chain, plotted as the gray dashed-dotted line. This is the case of degenerate (invariant) chains, where the closed system scenario is known to be more efficient than Markovian dephasing \cite{PlenioTransport}.

First, one can see from Fig. \ref{Coupling23} that transport efficiency decays with the chain size. 
This is expected since the number of local dissipators also increases.
It is remarkable that the presence of non-Markovianity leads to a significant enhancement of transport efficiency even for larger chains.
More interestingly,  transport efficiency with non-Markovian dephasing, $\eta_{\text{NM}}$, is larger than in the absence of dephasing, $\eta_{\text{ND}}$, and with Markovian dephasing only, $\eta_\text{{M}}$. This is a central result in our article, since it adds to and expands important previous results which compared Markovian and closed system dynamics \cite{PlenioTransport}.

\begin{figure}[h!]
\centering % para centralizarmos a figura
\includegraphics[width=7.2cm]{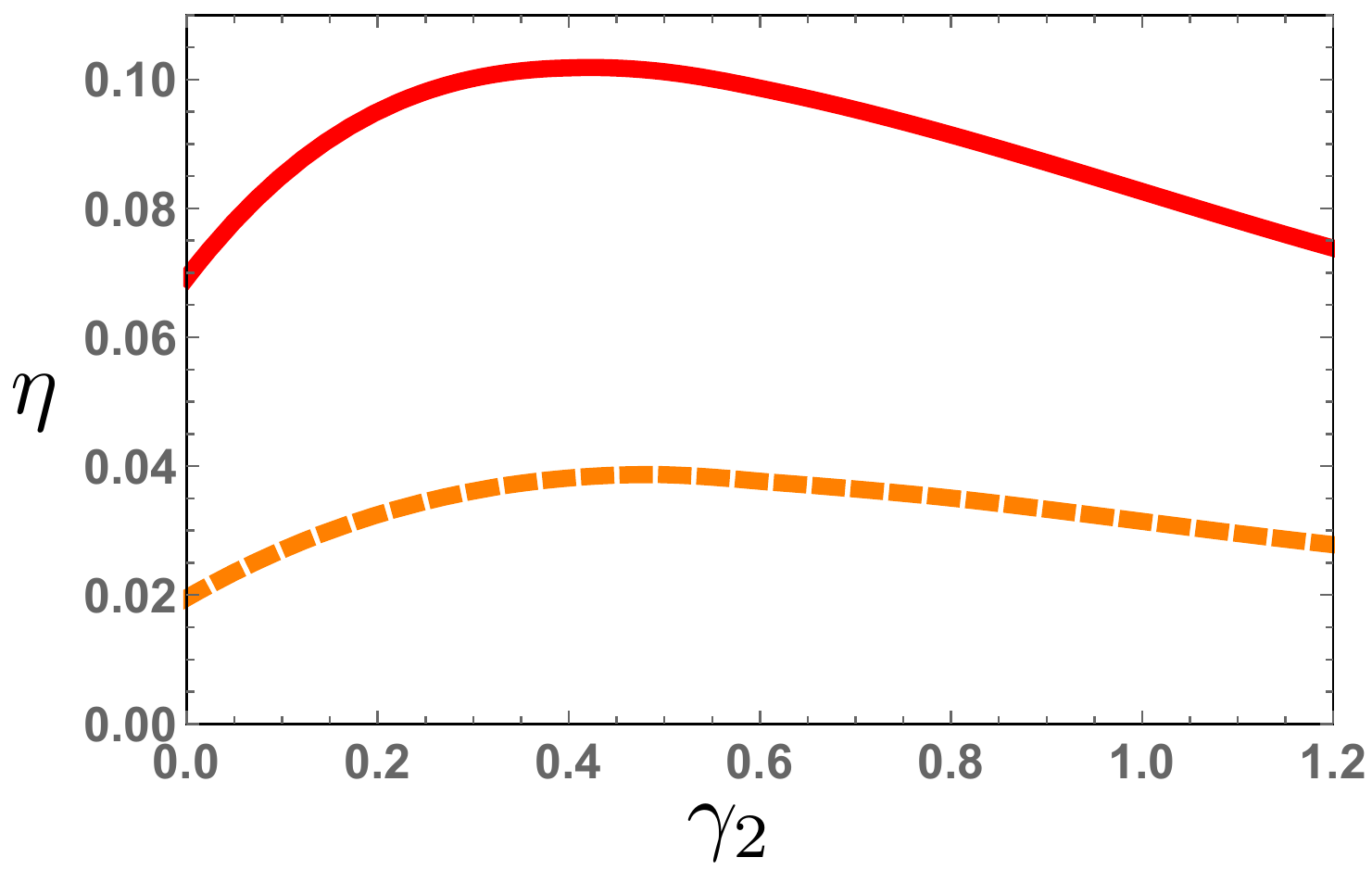}
\label{fig:fig14}
\caption{Plots of the transport efficiency $\eta$ as a function of $\gamma_2$ for a chain with $N=3$.  The values of the parameters are $\gamma_1(t)=\gamma_3(t)=0$,  $\kappa_1=\kappa_2=\kappa_3=0.05$, $\omega_1=\omega_2/4=\omega_3=0.5$, $\lambda_1=\lambda_2=0.2$, and $\kappa_{\text{sink}}=0.6$. The  dashed orange line corresponds to the Markovian case, while the red solid line corresponds to the non-Markovian scenario ($J=10,\theta=0.8$).
 } \label{DephasingAssistedT}
\end{figure}

It is important to note that we assumed only local dissipation and dephasing, which is usually the first model attempted when discussing new phenomena. It is under such models, for instance, that dephasing-assisted transport has been discovered \cite{PlenioTransport, Guzik}. At the same time, recent works have been investigating the limits of validity of the local assumption \cite{Gonzalez, Hofer, DeChiara, Mitchison, McConnella, Santos}.
For example, in Ref. \cite{Gonzalez} it is found that for weak intercoupling strength between the sites of a harmonic chain, a local master equation gives the correct stationary state when compared to the exact solution.
As the intercoupling strength becomes comparable to the local frequencies, the local master equation tends to not give the correct stationary state.
Therefore, energy transport is correctly described by a local master equation in this regime, while for strong intercoupling strength, a global master equation is found to be more suited to describe it.
We have seen that the enhancement in transport efficiency due to the presence of non-Markovianity in the model studied in this paper is significant for smaller $\lambda$ and tends to vanish as $\lambda$ is increased.
In this way, a question which deserves further investigation is whether a global master equation valid for strong intersite coupling $\lambda$ would enable more compelling results concerning the enhancement in transport efficiency.
For the non-Markovian model used, the local assumption follows the same reasoning. 
The model is quite accurate for small values of $\lambda$, which is the regime where the enhancement is more pronounced. 
A complete study taking into account correlated non-Markovian dephasing and discussing the dynamics for stronger intersite couplings will be shown elsewhere.

Finally, we revisit the phenomenon of dephasing-assisted transport for the model considered but now from the point of view of non-Markovianity. Under certain conditions, including position-dependent site energies, it is known that the transport efficiency is optimized when the chain is subjected to non-null dephasing \cite{PlenioTransport,Guzik}. 
 Consequently, closed system dynamics is not always the best choice to achieve high transport efficiency. 
 Recent investigations have indicated that Markovian dephasing-assisted transport arises from two competing processes, which are the tendency of dephasing to make the exciton population uniform and the formation of an exciton density gradient, defined by the source and the sink \citep{Elinor, Dubi}.
 
 In  Fig. \ref{DephasingAssistedT}, the parameters of the chain are chosen such that the manifestation of dephasing-assisted transport is possible, as in Ref. [9].  In this way, the parameters of the chain, $\kappa_i$, $\gamma_i(t)$ and $\omega_i$, where $i=1,2,3$ refers to the chain sites, are $\gamma_1(t)=\gamma_3(t)=0$,  $\kappa_1=\kappa_2=\kappa_3=0.05$, $\omega_1=\omega_2/4=\omega_3=0.5$, $\lambda_1=\lambda_2=0.2$, and $\kappa_{\text{sink}}=0.6$.
We then plotted the efficiency $\eta$ as a function of $\gamma_2$, both in the Markovian case, in dashed-dotted orange, and non-Markovian case, in red solid line.
%Notice that the only source of non-Markovian dephasing for the transport efficiency plotted in red corresponds to $\gamma_2(t)<0$ for some $t$, which occurs for all points in the red line since $J=10$ and $\theta=0.8$ in \eqref{gammat}.
%For the latter, besides $\gamma_1(t)=\gamma_3=0(t)$, we also set $J=10$ and $\theta=0.8$ in  \eqref{gammat}. 
One can see that in both cases the efficiency is maximized for non-null dephasing.
Once again, it is higher in the non-Markovian case than for the Markovian counterpart of the dynamics.

Besides that, the difference between the maximum of the efficiency $\eta$ and the value $\eta(\gamma_2=0)$ is $0.0327$ for the solid red line, while it is only $0.0192$ for the dashed-dotted orange line.
Therefore, we see that non-Markovianity can also boost efficiency in a scenario of dephasing-assisted transport.
%Therefore, we see that the presence of non-Markovian dephasing can lead to better results in assisting quantum transport than Markovian dephasing.
%In the context of controlled quantum systems, this possibility is particularly interesting, as one could set the values of the parameters such that non-Markovian dephasing can assist better quantum transport.
In the context of controlled quantum systems, this possibility is particularly interesting. In other words, one can then set the values of the parameters such that non-Markovianity may notably assist quantum transport.
Nonetheless, even though we have shown that dephasing-assisted transport can be more pronounced in the case of non-Markovian dephasing, future investigations on the conditions such that this happens, for other physical systems, would be potentially interesting.
%Therefore, we see that the phenomenon of dephasing assisted transport is more pronounced in the case of non-Markovian than in the case of Markovian dephasing.
%\section{Discussion}

 \section{Conclusion}\label{conc}

To summarize, we considered quantum engineered environments in the context of quantum transport. 
%For the model used, non-Markovianity led to the enhancement of energy transport efficiency in a dissipative chain.  
For the non-Markovian dephasing model used in this work, we observed larger transport efficiencies than in the case of Markovian dephasing and even in the absence of dephasing. The latter is observed for degenerate nearest neighbor coupled chains, a scenario well know to be optimum when compared to Markovian dephasing. 
In addition, we showed that dephasing-assisted transport also takes place in the non-Markovian case and that the efficiency of this mechanism is also improved by non-Markovianity. We hope that our work may motivate experiments in quantum transport using coupled quantum systems subjected to engineered environments. At the same time, further theoretical investigations using other schemes to control non-Markovianity, as well as approaches with global master equations valid for strong intercoupling strength, may reveal optimized strategies to minimize dissipation while achieving higher transport efficiency. Finally, although our study focuses on controlled systems, it sheds light on how quantum transport can benefit from non-Markovianity. 
We expect, therefore, that it will motivate the study of non-Markovianity in more complex quantum transport scenarios, controlled or otherwise.

\textit{Acknowledgements} -- S.V.M. acknowledges support from the Brazilian agency CAPES. F. L. S.
acknowledges partial support from of the Brazilian National
Institute of Science and Technology of Quantum Information
(CNPq INCT-IQ 465469/2014-0) and CNPq (Grant No. 302900/2017-9). D.O.S.P. acknowledges the Brazilian funding agencies CNPq (Grants No. 305201/2016-6), 
FAPESP (Grant No. 2017/03727-0) and the Brazilian National Institute of Science and Technology of Quantum Information (INCT/IQ).
BM, RRP, LSC and FLS acknowledge support from CAPES under CAPES/PrInt - process no. 88881.310346/2018-01.
All the authors would like to thank the SPIN off QuBIT research network for facilitating collaboration between researchers in the state of S\~ao Paulo.


\begin{thebibliography}{99}

\bibitem{Horodecki} R. Horodecki, P. Horodecki, M. Horodecki, and K. Horodecki, Rev. Mod. Phys. {\bf 81}, 865 (2009).
\bibitem{Baumgratz} T. Baumgratz, M, Cramer and M. B. Plenio, Phys. Rev. Lett. {\bf 113}, 140401(2014).
\bibitem{Moreira4} S. V. Moreira and M. Terra Cunha, Phys. Rev. A {\bf 99}, 022124 (2019).
\bibitem{Budini} A. Budini, Phys. Rev. Lett, {\bf 121}, 240401 (2018).
\bibitem{RevMarkov} A. Rivas, S. F. Huelga and M. B. Plenio, Rep. Prog. Phys. \textbf{77}, 094001 (2014).
\bibitem{Li} L. Li, M. J.W. Hall, and H. M. Wiseman, Phys. Rep. {\bf 759}, 1 (2018).
\bibitem{Breuer} H.-P. Breuer, E.-M. Laine, J. Piilo, and B. Vacchini, Rev. Mod. Phys {\bf 88}, 021002 (2016).
\bibitem{deVega} I. de Vega and D. Alonso, Rev. Mod. Phys. {\bf 89}, 015001 (2017).
\bibitem{PlenioTransport} M. B. Plenio, and S. F. Huelga, New J. Phys. {\bf 10} 113019 (2008).
\bibitem{Guzik} P. Rebentrost, M. Mohseni, L. Kassal, S. Lloyd, and A. Aspuru-Guzik, New J. Phys. \textbf{11} 033003 (2009).
\bibitem{poly} E. Collini and Gregory D. Scholes, Science \textbf{323}, 369 (2009).
\bibitem{Biggerstaff} D. N. Biggerstaff, R. Heilmann, A. A. Zecevik, M. Grafe, M. A. Broome, A. Fedrizzi, S. Nolte, A. Szameit, A. G. White and I. Kassal, Nat. Comm. {\bf 7}, 11282 (2015).
\bibitem{Schachenmayer} J. Schachenmayer, C. Genes, E. Tignone, and G. Pupillo, Phys. Rev. Lett. {\bf 114}, 196403 (2015).
\bibitem{Semiao} F. L. Semi\~ao, K. Furuya, and G. J. Milburn,  N. J. Phys. {\bf 12} 083033 (2010).
\bibitem{Feist} J. Feist and F. J. Garcia-Vidal, Phys. Rev. Lett. {\bf 114}, 196402 (2015).
\bibitem{EngVibAssEnTranfIons} D.J. Gorman, B. Hemmerling, E. Megidish, S. A. Moeller, P. Schindler, M. Sarovar, and H. Haeffner, Phys. Rev. X. \textbf{8}, 011038 (2018).
\bibitem{fmo1} A.W. Chin, J. Prior, R. Rosenbach, F. Caycedo-Soler, S.F.
Huelga, and M. B. Plenio, Nat. Phys. \textbf{9}, 113 (2013).
\bibitem{fmo2} P. Rebentrost, R. Chakraborty, and A. Aspuru-Guzik, J.  Chem. Phys. \textbf{131}, 184102 (2009).
\bibitem{fmo3} S. Jang, S. Hoyer, G. Fleming, and K. Birgitta Whaley, Phys. Rev. Lett. \textbf{113}, 188102 (2014).
\bibitem{fermi1} R. Aguado and T. Brandes, Phys. Rev. Lett. \textbf{92}, 206601 (2004).
\bibitem{fermi2} P. Ribeiro and V. R. Vieira, Phys. Rev. B \textbf{92}, 100302 (2015).
\bibitem{fermi3} P. Zedler, G. Schaller, G. Kiesslich, C. Emary, and T. Brandes, Phys. Rev. B \textbf{80}, 045309 (2009).
\bibitem{EnvoiAssTranspIonsMaier} C. Maier,  T. Brydges, P. Jurcevic, N. Trautmann, C. Hempel, B. P. Lanyon, P. Hauke, R. Blatt, and C. Roos, Phys. Rev. Lett.  \textbf{122}, 050501 (2019).
\bibitem{paulo} P. C. C\'ardenas, M. Paternostro, and F. L. Semi\~ao, Phys. Rev. A \textbf{91}, 022122 (2015).
\bibitem{Souza} A. M. Souza, J. Li, D. O. Soares-Pinto, R. S. Sarthour, I. S. Oliveira, S. F. Huelga, M. Paternostro, and F. L. Semi\~ao, arXiv:1308.5761.
\bibitem{fred} F. Brito and T. Welang, New J. Phys. \textbf{17}, 072001 (2015).
\bibitem{mauro} S. Lorenzo, F. Plastina, and Mauro Paternostro, Phys. Rev. A, \textbf{87}, 022317 (2013).
\bibitem{exp}  F. Wang, P. -Y. Hou, Y. -Y. Huang, W. -G. Zhang, X. -L. Ouyang, X. Wang, X. -Z. Huang, H. -L. Zhang, L. He, X. -Y. Chang, and L. -M. Duan, Phys. Rev. B \textbf{98}, 064306 (2018).
\bibitem{nmqkd} R. Vasile, S. Olivares, M. G. A. Paris, and S. Maniscalco, Phys. Rev. A \textbf{83}, 042321 (2011).
\bibitem{nmqm} A. W. Chin, S. F. Huelga, and M. B. Plenio, Phys. Rev. Lett. \textbf{109}, 233601 (2012).
\bibitem{nmqt} E. -M. Laine, H. -P. Breuer, and Jyrki Piilo, Sci. Rep. \textbf{4}, 4620 (2014).
\bibitem{BLP} H. -P.  Breuer, E. -M. Laine, and J. Piilo, Phys. Rev. Lett. \textbf{103}, 210401 (2009).
\bibitem{RHP} A. Rivas and S. F. Huelga,  Phys. Rev. Lett. \textbf{105}, 050403 (2010).
\bibitem{Pollock} F. A. Pollock, C. Rodr\'iguez-Rosario, T. Frauenheim, M. Paternostro, and K. Modi, Phys. Rev. A {\bf 97}, 012127 (2018).
\bibitem{Pollock2} F. A. Pollock, C. Rodr\'iguez-Rosario, T. Frauenheim, M. Paternostro, and K. Modi, Phys. Rev. Lett. {\bf 120}, 040405 (2018).
\bibitem{Milz} S. Milz, M. S. Kim, Felix A. Pollock, and K. Modi, Phys. Rev. Lett. {\bf 123}, 040401 (2019).
\bibitem{Buscemi} F. Buscemi and N. Datta, Phys. Rev. A {\bf 93}, 012101 (2016).
%\bibitem{Moreira5} S. V. Moreira and F. L. Semi\~ao, arXiv: 1808.04427 (2018).
\bibitem{Gonzalez} J. O. Gonz\'alez, L. A. Correa, G. Nocerino, J. P. Palao, D. Alonso and G. Adesso, Open Syst. Inf. Dyn. {\bf 24}, 1740010 (2017).
\bibitem{Hofer} P. P. Hofer, M. Perarnau-Llobet, L. M. Miranda, G. Haack, R. Silva, J. B. Brask and N. Brunner,  New J. Phys. {\bf 19}, 123037 (2017).
\bibitem{DeChiara} G. De Chiara, G. Landi, A. Hewgill, B. Reid, A. Ferraro, A. J. Roncaglia and M. Antezza, New J. Phys. {\bf 20}, 113024 (2018).
\bibitem{Mitchison}  M. T. Mitchison and M. B. Plenio, New J. Phys. {\bf 20}, 033005 (2018). 
\bibitem{McConnella} C. McConnella and  A. Nazirb, J. Chem. Phys. {\bf 151}, 054104 (2019).
\bibitem{Santos} J. P. Santos and F. L. Semi\~ao, Phys. Rev. A {\bf 89}, 022128 (2014).
\bibitem{Elinor} E. Zerah-Harush and Y. Dubi, J. Phys. Chem. Lett. {\bf 9}, 1689 (2018).
\bibitem{Dubi} Y. Dubi, J. Phys. Chem. C {\bf 119}, 25252 (2015).




%\bibitem{Emary2} C. Emary, Phys. Rev. A {\bf 96}, 042102 (2017).
%\bibitem{Hamm} P. Hamm. Principles of nonlinear optical spectroscopy: A practical approach.
%\bibitem{Baumgratz} T. Baumgratz, M, Cramer and M. B. Plenio, Phys. Rev. Lett. {\bf 113}, 140401(2014).
%\bibitem{Streltstov} A. Streltstov, G. Adesso and M. B. Plenio, Rev. Mod. Phys. {\bf 89}, 041003 (2017).
%\bibitem{Meier} C. Meier, D. J. Tannor, J. Chem. Phys. {\bf 111} (1999) 3365.
%\bibitem{Dostal} J. Dostal, B. Benesova and T. Brixner,  J. Chem. Phys. {\bf 145}, 124312 (2016).
%\bibitem{Kurnit} N. A. Kurnit, I. D. Abella, and S. R. Hartmann, Phys. Rev. Lett. {\bf 13}, 567 (1964).
%\bibitem{Leggett} A. J. Leggett, Science, {\bf 307}, 871 (2005).
%\bibitem{LeggettM} A. J. Leggett, P. Theor. Phys. Sup. {\bf 69}, 80 (1980).
%\bibitem{Leggett3} A. J. Leggett, J. Phys. Condens. Matt. {\bf 14}, R415 (2002).
%\bibitem{Wilde} M. M. Wilde and A. Mizel, Found. of Phys. {\bf 42}, 256 (2012).
%\bibitem{Jonas} D. M. Jonas, Annu. Rev. Phys. Chem {\bf 54}, 425 (2003).






\end{thebibliography}
\end{document}